\documentclass[mathleft]{an}
\usepackage{graphicx}
\usepackage{times}
\begin{document}

% The following seven commands are intended for editorial usage and should be ignored by
% the author(s).
\Pagespan{789}{}% Document's page range. 
% If second parameter is left empty, the last page is computed automatically.
\Yearpublication{2008}%
\Yearsubmission{2008}%
\Month{11}%   
\Volume{999}%  
\Issue{88}% 
% \DOI{This.is/not.aDOI}% 

\title{High M/L ratios of UCDs: a variation of the IMF?}

\author{S. Mieske\inst{1,2}\fnmsep\thanks{Corresponding author:
  \email{smieske@eso.org}\newline}
%Example 
%for footnote, note the usage of the \texttt{fnmsep}
%command as separator between institute number and footnote mark} 
\and  J. Dabringhausen\inst{3} \and P. Kroupa\inst{3} \and M. Hilker\inst{2} \and H. Baumgardt\inst{3}
}
\titlerunning{IMF variation in UCDs?}
\authorrunning{S. Mieske et al.}
\institute{
European Southern Observatory, Alonso de Cordova 3107, Vitacura, Santiago, Chile
\and 
European Southern Observatory, Karl-Schwarzschild-Str.2, 85748 Garching b. Muenchen, Germany
\and 
Argelander Institut f\"ur Astronomie, Auf dem H\"ugel 71, 53121 Bonn, Germany}

\received{22 August 2008}
\accepted{tbd}
\publonline{tbd}


\abstract{Various studies have established that the dynamical M/L
  ratios of ultra-compact dwarf galaxies (UCDs) tend to be at the
  limit or beyond the range explicable by standard stellar populations
  with canonical IMF. We discuss how IMF variations may account for
  these high M/L ratios and how observational approaches may in the
  future allow to discriminate between those possibilities. We also
  briefly discuss the possibility of dark matter in UCDs. }

\maketitle

\section{Introduction: elevated M/L ratios in UCDs}
\begin{figure}
\includegraphics[width=83mm]{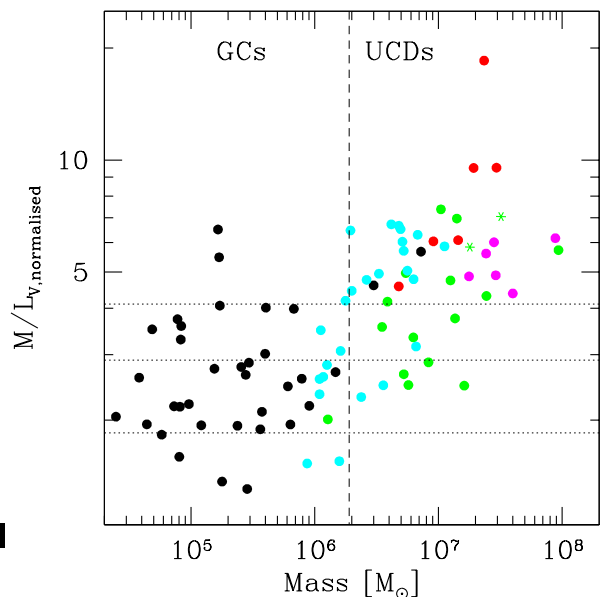}
\caption{Mass vs. M/L for compact stellar systems. Literature sources
are Mieske et al. (2008, green dots), Rejkuba et al. (2007, cyan),
Ha\c{s}egan et al. (2005, red) Evstigneeva et al. (2007, magenta),
Hilker et al. (2007, green asterisk), and Milky Way globular clusters
(McLaughlin \& van der Marel 2005, black dots).  The vertical dashed
line at 2$\times 10^6 M_{\odot}$ indicates the approximate mass where
the relaxation time is equal to one Hubble time (Mieske et al. 2008;
Dabringhausen et al. 2008). All M/L ratio estimates have been
normalised to solar metallicity (Mieske et al. 2008).  The horizontal
dotted lines indicate the M/L ratios expected for single stellar
populations of age 13, 9, and 5 Gyrs (from top to bottom, based on
Bruzual \& Charlot 2003 and Maraston et al. 2005).}
\label{fig1}
\end{figure}

In recent years, significant effort has been put into studying the
internal dynamics of extragalactic compact stellar systems in the mass
regime of massive globular clusters and ultra-compact dwarf galaxies
($10^6<M/M_{\sun}<10^8$) (Drinkwater et al. 2003, Martini \& Ho 2004,
Ha\c{s}egan et al. 2005, Maraston et al. 2004, Rejkuba et al. 2007,
Evstigneeva et al. 2007, Hilker et al. 2007, Mieske et al. 2008; see
also Dabringhausen et al. 2008 and Forbes et al. 2008). A striking
outcome of these studies is that the dynamical M/L ratios of massive
compact stellar systems are on average about two times larger than for
normal globular clusters of comparable metallicity. This is
illustrated in Fig.~\ref{fig1}. The rise of M/L ratios starts at about
2$\times 10^6 M_{\odot}$, separating ordinary globular clusters at
lower masses, from the so-called ultra-compact dwarf galaxies, at
higher masses (see also Mieske et al. 2008). Possible reasons for
these high M/L ratios include extreme stellar mass functions (Mieske
\& Kroupa 2008; Dabringhausen, Hilker \& Kroupa 2008) or densely
packed dark matter (Goerdt et al. 2008).

\section{Possible explanations}
\subsection{Bottom-heavy IMF}
A bottom heavy stellar initial mass function (IMF) may be a possible
explanation for the high M/L$_V$ ratios of UCDs. Such steep low-mass
stellar IMFs may result if the radiation field, stellar winds and
supernova explosions in UCDs were so intense during their formation,
that pre-stellar cloud cores were ablated before they could fully
condense to stars (Kroupa \& Bouvier 2003).  By confirming such an
overabundance of low-mass stars with respect to a canonical IMF
(Kroupa 2001), we would for the very first time have unambiguous
evidence for a radically different star-formation process under
extreme physical conditions when UCDs formed.

\begin{figure*}
\includegraphics[width=53mm]{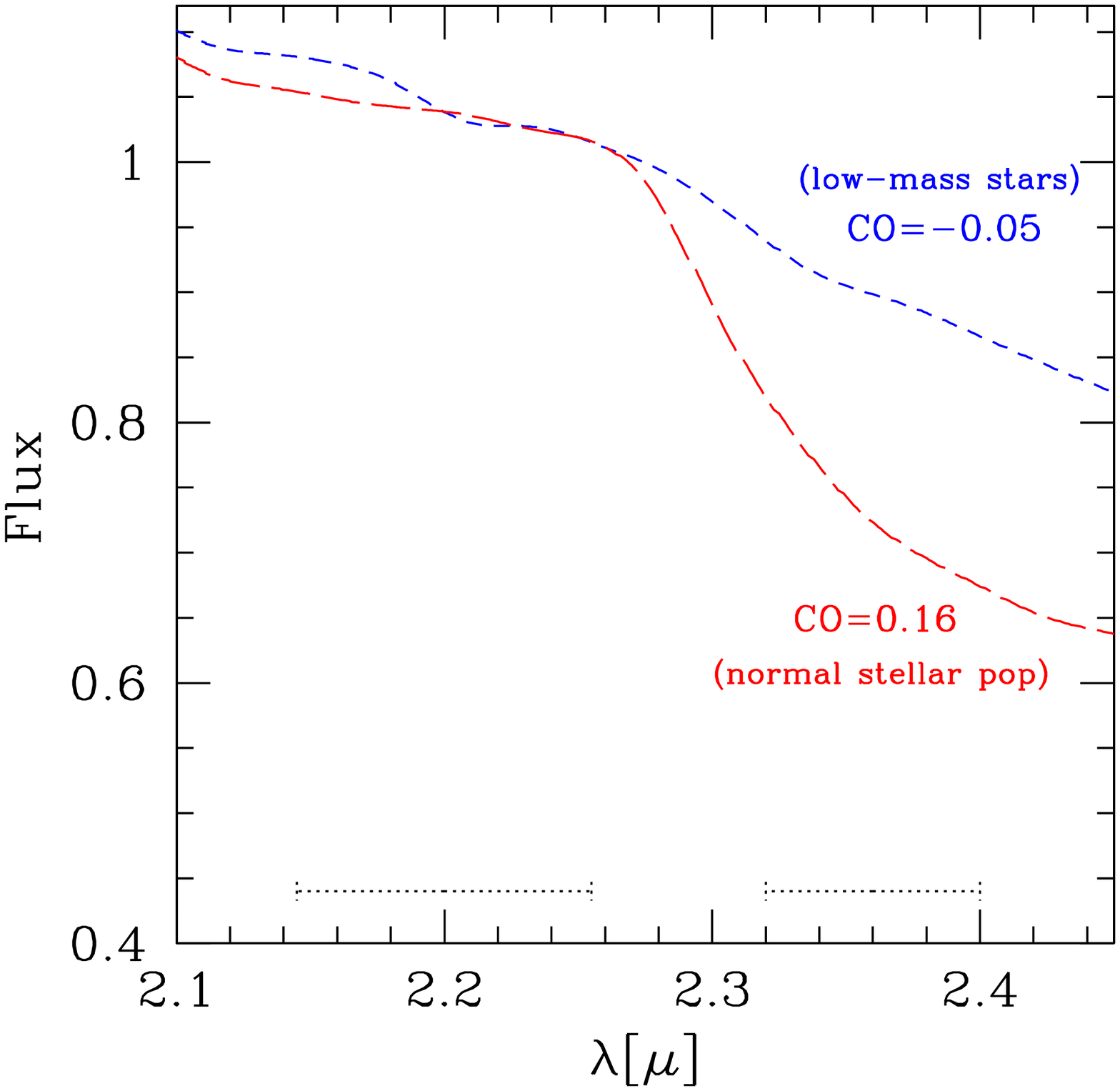}\hspace{0.3cm}
\includegraphics[width=53mm]{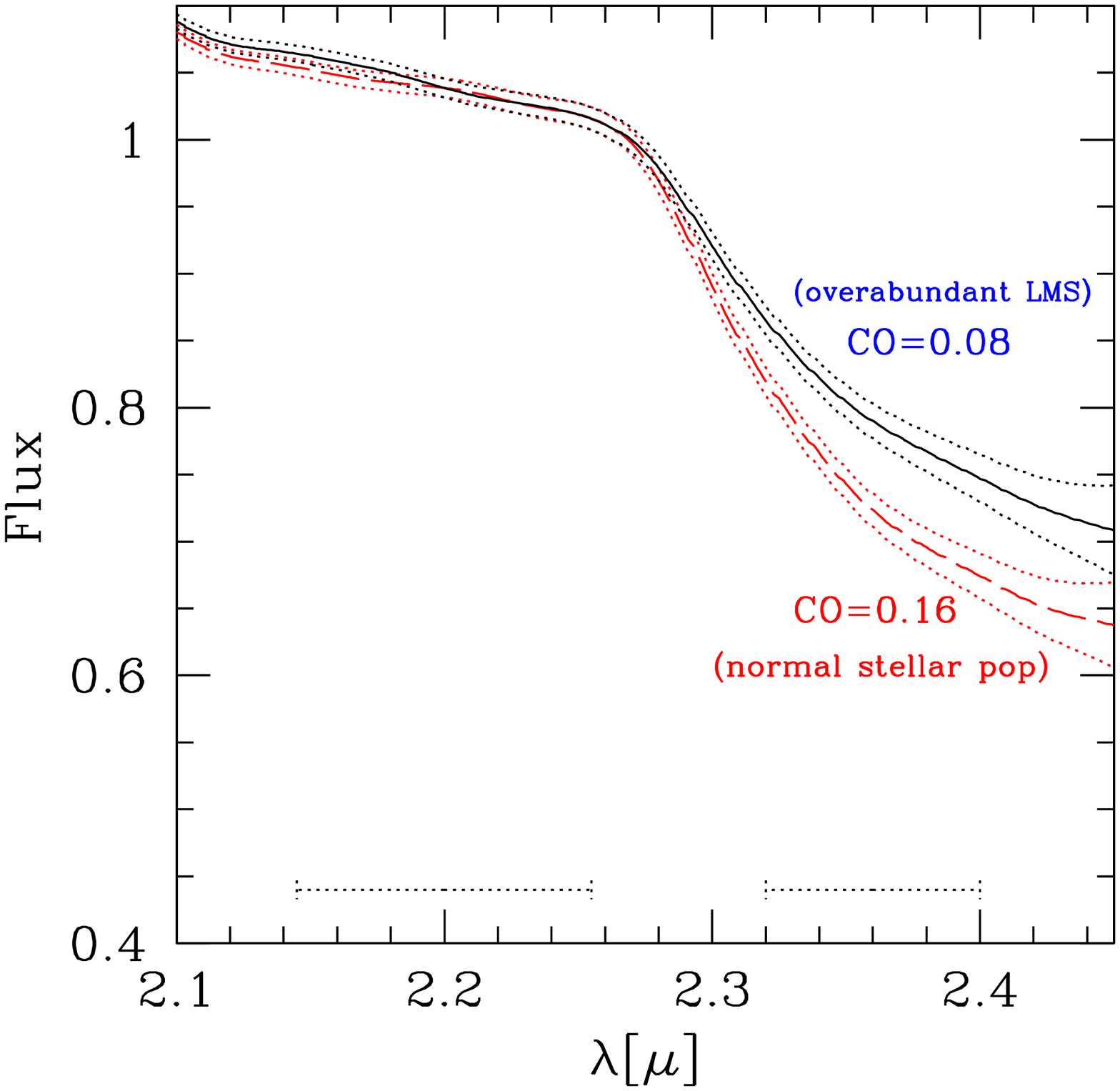}\hspace{0.3cm}
\includegraphics[width=58mm]{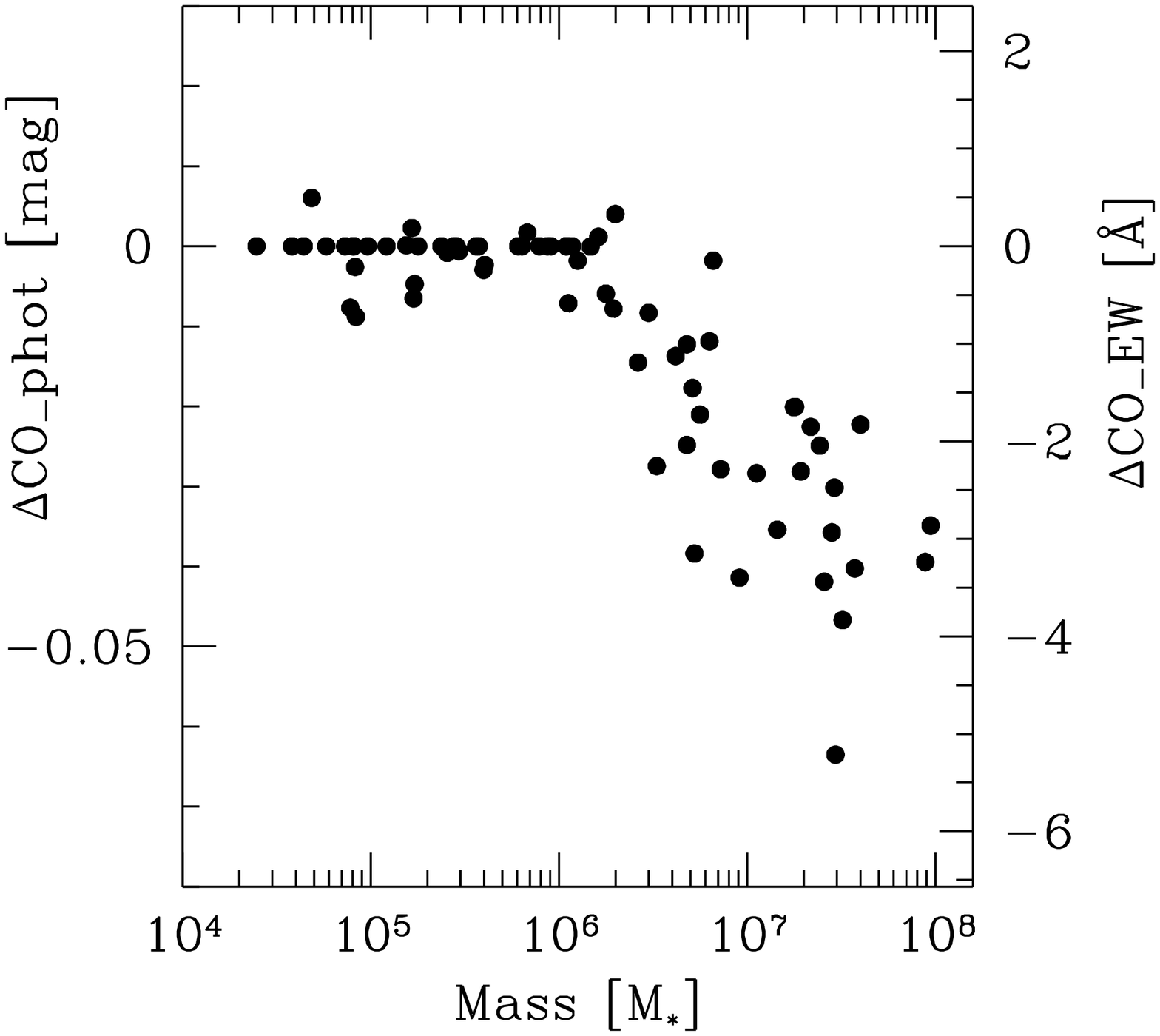}

\caption{These plots illustrate how an overabundance of low-mass stars
    in an old stellar population can be estimated from spectroscopy
    around the CO band (Kroupa \& Gilmore 1994). The feature and
    continuum band definition of the photometric index is indicated by
    dotted lines at the bottom (Frogel et al. 1978). See also Mieske
    \& Kroupa 2008 for more details. {\bf Left:} Shown is a NIR
    spectrum of a M2II giant star (lower long dashed curve) and a M2V
    dwarf star (upper short dashed curve) from the catalog of Lancon
    \& Rocca-Volmerange (1992), smoothed to 0.04$\mu$ resolution.  The
    CO index 0.16 mag of the M giant corresponds to the typical CO
    index of old stellar populations of intermediate metallicity
    ([Fe/H]$\sim-$0.7 dex, e.g.  Frogel et al. 1978 and 2001, Goldader
    et al.  1997, Ivanov et al.  2000).  The M dwarf has a much weaker
    CO feature, representative of a population of pure low-mass stars
    with upper mass cutoff $m_{\rm cut}\simeq 0.5$ M$_{\odot}$
    (Kroupa \& Gilmore 1994).  Under the hypothesis that the high M/L
    ratios of UCDs are caused by an overabundance of unevolved
    low-mass stars like the M dwarf (Kroupa \& Gilmore 1994), one
    would require those stars to contribute a significant fraction to
    the total UCD mass (3/4 for the highest M/L UCD known), and also a
    certain fraction to the K-band luminosity (see next panel).  {\bf
    Middle:} The spectrum of the canonical stellar population is shown
    as in the left panel. The solid line now indicates the case of
    40\% K-band luminosity contribution from an additional population
    of low-mass stars. This luminosity fraction is representative for
    the case when the additional low-mass stars make up 3/4 of the
    total mass, at a metallicity of about [Fe/H]=$-$0.7 dex (Mieske \&
    Kroupa 2008). The dotted lines indicate the 1 $\sigma$ error range
    of the spectrum, assuming a S/N of 150 per 0.04$\mu$ resolution
    element. {\bf Right:} The expected offset from the Frogel et
    al. [Fe/H]$-$CO relation (Frogel et al. 2001) is plotted vs. the
    mass of the compact stellar systems from Fig.~\ref{fig1}. The
    offset arises from the assumption that M/L ratios above the mean
    value for galactic GCs are caused by a bottom-heavy IMF.}
\label{fig2}
\end{figure*}

For diagnosing a bottom heavy IMF, one needs to study a portion of the
spectrum where the hypothetical overabundant population of low-mass
(main sequence) stars contributes significantly to the integrated
spectrum. This is the case for the near-infrared wavelength region of
the CO band ($2.3 \mu < \lambda < 2.42 \mu$, see Fig.~\ref{fig2}).
For intermediate to high metallicities, late type giant stars have a
very strong CO absorption feature, while late type dwarf stars have a
much weaker feature (see also Mieske \& Kroupa 2008), independent of
metallicity (e.g. Frogel et al.  1978).  At a given metallicity, one
thus expects a weaker CO index for high M/L ratio sources, if the high
M/L ratios are caused by low-mass stars. This is quantified in
Fig.~\ref{fig2}, and derived in more detail in Mieske \& Kroupa
(2008). We have embarked on an observational study with ISAAC@VLT to
compare the CO index depth of the two UCDs with highest M/L ratios to
values found for globular clusters with ``normal'' M/L (ESO observing
period 81 and 82). We expect a 3-4 $\sigma$ signal if the elevated M/L
is caused by a bottom-heavy IMF.

\subsection{Top heavy IMF}
\begin{figure*}[h!]
\begin{center}
\includegraphics[width=8.2cm]{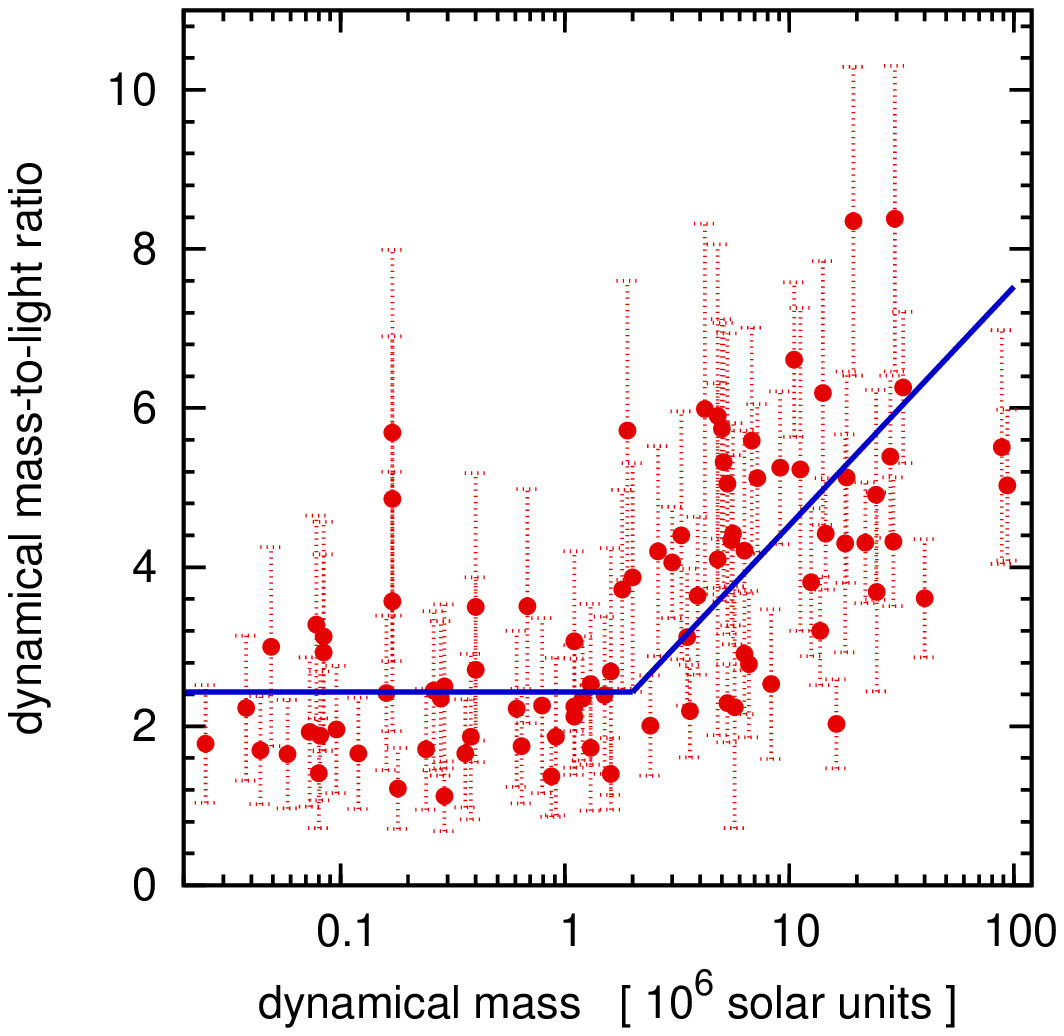}\hspace{0.3cm}
\includegraphics[width=8.2cm]{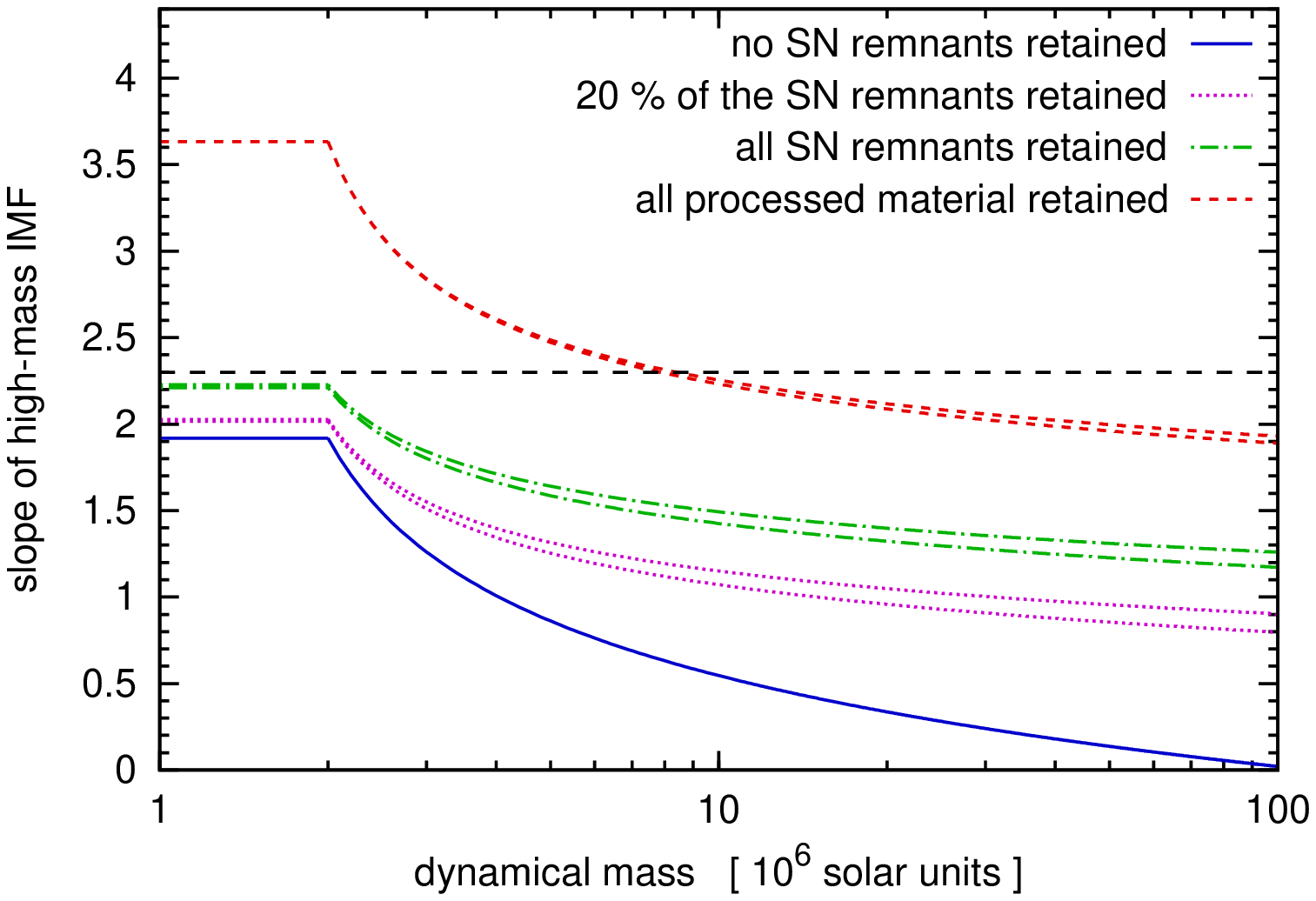}

\end{center}
\caption{{\bf Left:} Data points from Fig.~\ref{fig1}, but with linear
Y-axis. The (blue) solid line for masses below $2 \times 10^6
M_{\odot}$ indicates the mean M/L ratio of the sources. For larger
masses, the (blue) solid line indicates a linear fit to the data
points. {\bf Right:} Assuming the behaviour of M/L as f (M) as
indicated by the solid lines in the left panel, this figure indicates
how the high-mass IMF slope above 1 solar mass changes as f(M). The
various line shapes correspond to different cases of retained
remnants, as indicated in the plot legend. The horizontal dashed line
indicates the canonical IMF slope (Kroupa 2001).}
\label{fig3}
\end{figure*}

As opposed to a bottom-heavy IMF, an elevated M/L ratio may also be
caused by a large number of stellar remnants, i.e. an originally
top-heavy IMF. This seems attractive, since it is also suggested by
models for galaxy evolution (e.g. Baugh et al. 2005; Nagashima et
al. 2005; van Dokkum 2008) or GC evolution (e.g. D'Antona \& Caloi
2004; Prantzos \& Charbonnel 2006). In Dabringhausen, Kroupa \&
Baumgardt (2008, submitted to MNRAS), it is examined in detail how the
top-heavy IMF slope $\alpha_3$ changes as M/L ratio increases. For
this, they assume the IMF to be a multi-power law equal to the
canonical IMF (e.g.  Kroupa 2001) below $1 \, M_{\odot}$ but with a
different slope, $\alpha_3$, above $1 \, M_{\odot}$. Their results are
indicated in Fig.~\ref{fig3}. For the parametrisation of $\alpha_3$ as
a function of the mass $M$ (expressed in solar masses) of a stellar
system (right panel of Fig.~\ref{fig3}), we suggest
\begin{equation}
\overline{\alpha_3}(M)=\log ( \frac{0.47 (10^{-6} \, 
M)^2}{(10^{-6} \, M)-1})^{-1}+0.41
\label{eqalpha3}
\end{equation}
for $M \ge 2 \times 10^6 \, {M}_{\odot}$, and
$\overline{\alpha_3}(M)=2.3$ for $M < 2 \times 10^6 \,
{M}_{\odot}$.  $\overline{\alpha_3}(M)$ thus returns the
canonical IMF for GCs, which is motivated with the invariance of the
IMF in resolved stellar populations (e.g. Kroupa 2001).

In order to test the hypothesis of a shallower IMF at high masses, one
needs to determine the frequency of high-mass stellar remnants as a
function of M/L in UCDs and GCs. A top-heavy IMF implies an
over-abundance of neutron stars / black holes, and implicitly and
overabundance of low-mass X-ray binaries (LMXBs; Jordan et
al. 2004). Therefore, quantifiying the association probability of UCDs
and GCs with LMXBs as a function of their structural parameters and
their M/L ratio is the next step in terms of observational tests for
the origin of high M/L in UCDs.

\subsection{Dark matter}

\begin{figure}[h!]
\begin{center}
\includegraphics[width=8.3cm]{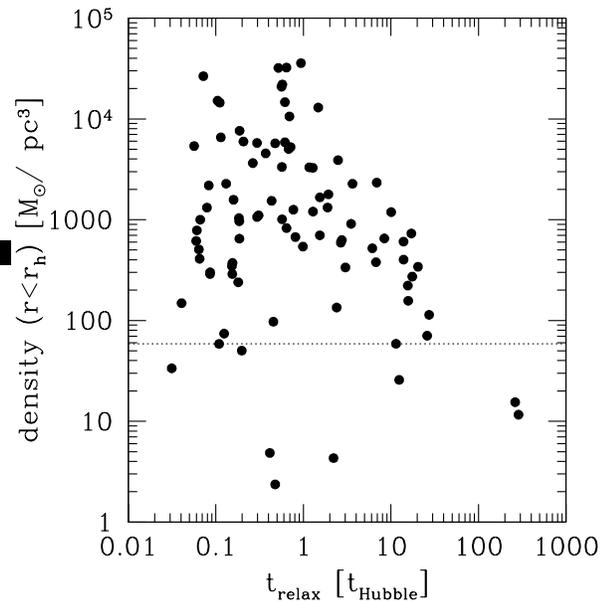}
\end{center}
\caption{The mean mass density within the half-mass radius of the
joint sample of GCs and UCDs from Fig.~\ref{fig1} is plotted
vs. their relaxation time (Mieske et al. 2008). The dotted line indicates
the approximate central ($r\leq 10$pc) dark matter densities
expected for cuspy dwarf galaxy CDM halos (Gilmore et al. 2007).}
\label{fig4}
\end{figure}

Goerdt et al. (2008) have shown that funneling of dark matter to the
central region of a disk galaxy, due to gas-infall, can significantly
increase the M/L ratios in the nuclear region, and hence may explain
the elevated M/L ratios of UCDs, provided that UCDs formed by tidal
stripping. Indeed, it has been suggested that also GCs may have
originated as centers of individual primordial dark matter halos
(e.g. Lee et al. 2007, Bekki et al.  2007). If dark matter funneling
is an efficient mechanism (Goerdt et al. 2008), one may therefore
expect both UCDs and GCs to be formed with a significant fraction of
dark matter. It is important to note that such an increase of dark
matter density by some kind of funneling mechanism is necessary to
explain a significant amount of dark matter in UCDs or GCs, since
their present-day stellar (and hence implied dark matter) densities
are up to 2-3 orders of magnitude higher than expected for cuspy dark
matter halos of dwarf galaxy mass (Gilmore et al. 2007). This is shown
in Fig.~\ref{fig4}. In Baumgardt \& Mieske (2008) the dynamical
co-evolution of stars and dark matter in GCs/UCDs is studied.

\acknowledgements

We thank the conference organisers of GSD08 in Strassburg for a very enjoyable meeting.

\newpage%%%%%%%%%%%%%%%%%%%%%%%%%%%%%%%%%%%%%%%%%%%%%%%%%%%%%%

\end{document}